\documentclass[a4paper,noarxiv,twocolumn,accepted=2022-03-20]{quantumarticle}
\usepackage[utf8]{inputenc}
\usepackage[english]{babel}
\usepackage[T1]{fontenc}
\usepackage[ruled,vlined]{algorithm2e}

\usepackage{graphicx}
\usepackage{amsmath}
\usepackage{bbold}

\usepackage{hyperref}
\usepackage{breakurl}

\title{Approaching the theoretical limit in quantum gate decomposition}

\author{Péter Rakyta}
\affiliation{Department of Physics of Complex Systems, E\"otv\"os  Lor\'and  University, Budapest, Hungary}
\affiliation{Wigner Research Center for Physics, 29–33 Konkoly–Thege Miklos Str., H-
1121 Budapest, Hungary}

\author{Zoltán Zimborás}
\affiliation{Wigner Research Center for Physics, 29–33 Konkoly–Thege Miklos Str., H-1121 Budapest, Hungary}
\affiliation{BME-MTA Lendület Quantum Information Theory Research Group, Budapest, Hungary}

\begin{document}

\maketitle

\begin{abstract}
In this work we propose a novel numerical approach to decompose general quantum programs in terms of single- and two-qubit quantum gates with a $CNOT$ gate count very close to the current theoretical lower bounds.
In particular, it turns out that $15$ and $63$ $CNOT$ gates are sufficient to decompose a general $3$- and $4$-qubit unitary, respectively, with   {high numerical accuracy}.
Our approach is based on a sequential optimization of parameters related to the single-qubit rotation gates involved in a pre-designed quantum circuit used for the decomposition.
In addition, the algorithm can be adopted to sparse inter-qubit connectivity architectures provided by current mid-scale quantum computers, needing only a few additional $CNOT$ gates to be implemented in the resulting quantum circuits.
\end{abstract}

\section{Introduction}

Quantum mechanical resources like superposition and entanglement are useful when solving computational problems that are intractable with classical resources, i.e., with classical computing devices. 
For example, Shor's integer factorization \cite{doi:10.1137/S0097539795293172} and Grover's database search \cite{PhysRevLett.79.325} algorithms promise a significant speedup compared to the best classical algorithms developed for these problems.
Even if currently available quantum computers are not advanced enough yet to outperform classical devices in solving these problems, a lot of progress has been made in demonstrating these algorithms on small quantum devices \cite{doi:10.1063/1.125846,Figgatt2017,Martin-Lopez2012,Monz1068,PhysRevA.100.012305}.
Beside these well known examples, various other promising schemes were developed to exploit quantum resources in solving computational problems, for example, variational quantum optimization \cite{harrigan2021quantum}, variational quantum eigensolvers   \cite{arute2020hartree}, and
quantum simulations of molecular and many-body phenomena \cite{Smith2019,Leontica2021,doi:10.1063/5.0046930, doi:10.1063/5.0040477, satzinger2021realizing} were successfully implemented on hardware, and at least qualitatively justified results were obtained.

In general, quantum algorithms can be described by unitary
transformations and projective measurements acting on the $2^n$-dimensional Hilbert space spanned by the computational basis states of $n$ quantum bits (qubits) involved in the program. 
The individual unitary transformations forming the quantum program are the quantum gates. A  widely used approach to measure the complexity of a quantum circuit (a quantum program) is in terms of the number of elementary gates involved in the circuit \cite{PhysRevA.52.3457} (or alternatively the depth of the circuit composed by these elementary gates).

In current quantum devices gates are still somewhat noisy
due to undesirable interactions with the environment.
Thus, by increasing the number of quantum gates in a circuit
makes it harder to retrieve any meaningful result from the quantum computer.
Therefore the ability to decompose quantum programs into low complexity gate arrays plays an important role in compiling programs onto currently available noisy quantum computers.
Also, we need to keep in mind that
some hardware components of a quantum computer is experimentally more demanding to realize than others.
Depending on the underlying architecture \cite{Linke3305} the operation error characteristic for one-qubit and two-qubit gates might differ even by an order of magnitude \cite{10.1145/3297858.3304007}.
Thus, it is a commonly used strategy of decomposition algorithms to focus only on making the number of two-gates as low as possible in the resulting quantum circuit.
Ref.~\cite{PhysRevA.69.062321} showed that the theoretical lower limit $N_{theo}$ of $CNOT$ gates sufficient to decompose any $n$-qubit unitary is given by 
\begin{equation}
   N_{theo}(n) = \lceil\tfrac{1}{4}\left(4^n - 3n - 1\right) \rceil \;. \label{eq:theo_limit}
\end{equation}
However, no circuit construction strategy has been so far developed that meets this lower limit of $CNOT$ gates.

A conventional, QR decomposition based approach \cite{PhysRevA.52.3457} to implement general multi-qubit gates in the decomposition yields an array of $O(n^34^n)$ elementary quantum gates.
Later the method was improved by employing Gray code basis \cite{1992nrfa.book.....P} and a decomposing algorithm requiring $8.7\cdot4^n$ $CNOT$ gates \cite{PhysRevLett.92.177902} was developed.
An analogous algorithm was developed by Shende et al \cite{1629135} improving the limit of $CNOT$ gates to $2\cdot4^n - (2n + 3)2n + 2n$ with iterative state preparation cycles to decompose a general unitary column by column.

Concurrently, an alternative method to decompose general unitaries was developed by the work of Refs.~\cite{tucci} and \cite{mottonen2005}.
The recursive algorithm based on the CSD method \cite{PAIGE1994303} makes it possible to decompose a general unitary with $\frac{1}{2}\cdot n\cdot 4^n -\frac{1}{2}\cdot 2^n$ $CNOT$ gates \cite{tucci}. 
By further optimizations this limit can be lowered to $\frac{1}{2}\cdot 4^n -\frac{1}{2}\cdot 2^n - 2$ $CNOT$ gates \cite{mottonen2005}.
The developed recursive CSD method is implemented in the Qubiter package \cite{qubiter_github} maintained up to these days.
Currently the most efficient decomposing algorithm was introduced in Ref.~\cite{1629135}.
The Quantum Shannon Decomposition (QSD) follows similar strategy than the recursive CSD method, just using eigenvalue decomposition during the recursive steps instead of the CSD method.
In each iteration step the decomposition is applied to unitary gates containing of one less qubit than in the previous step. 
The QSD needs $\frac{3}{4}\cdot 4^n - \frac{3}{2}\cdot 2^n$ $CNOT$ gates to decompose a general unitary, which limit can be  lowered to $\frac{23}{48}\cdot 4^n - \frac{3}{2}\cdot 2^n +\frac{4}{3}$ $CNOT$ gates by additional optimizations \cite{1629135}.
Recently an efficient implementation of the QSD method developed in the OpenQL \cite{khammassi2020openql} package was published in Ref.~\cite{krol2021efficient}.

Despite of the great progress in the field, the described decomposing algorithms suffer from a crucial limitation.
Namely, they assume full connectivity between the qubits during the decomposition.
However, to run quantum algorithms on currently available gate-model based quantum computers, quantum circuits must be adopted to the constraints of the underlying hardware.
Reference \cite{mottonen2005} reported on a modified CSD method adopted to an architecture with only nearest neighbour connections between the qubits.
Here the limitations of the connectivity were incorporated into the building blocks of the decomposing quantum circuit.
Using this modified algorithm the $CNOT$ gate count was increased by a factor of less than two compared to the case of full connectivity.
Another strategy to compile a quantum circuit on a realistic quantum device is to insert additional $SWAP$ operations to bring qubits close to each other to apply two-qubit controlled gates on them. 
To further decrease the gate count hardware specific strategies can be developed to determine the order of the individual $SWAP$ operations \cite{8491885, Sisodia2018CircuitOF}.
However, the resulting gate count seems to be still to large and versatile error-correction, mapping or scheduling measures need to be applied during the post processing of the quantum computational results \cite{Rotteler2008,khammassi2020openql,khammassi2020openql}.

In this work we propose a novel strategy to approximately decompose general unitaries into quantum circuits consisting of single- and two-qubit controlled quantum gates.
Our numerical method is based on an iterative, sequential optimization (SO) of the parameters of the incorporated quantum gates making it possible to support hardware specific limitations of the target device, without the need of any post-processing or addition of extra $SWAP$ gates.
In an ideal case of full connectivity between the qubits our algorithm requires $CNOT$ gate count very close to $N_{theo}(n)$ in order to decompose a general quantum program.
In particular, $15$ $CNOT$ gates turned to be sufficient to decompose a general $3$-qubit unitary with high numerical precision, while the decomposition of a $4$-qubit unitary requires only $63$ $CNOT$ gates compared to $20$ and $100$ $CNOT$ gates needed by the optimized QSD, the most efficient algorithm known today.
We note that all the previously mentioned algorithms -- including the SO method -- aim to capture the fewest possible number of $CNOT$ gates being sufficient to decompose any unitary. 
Besides this "worst case" scenario, the quest to find the most efficient decomposition of individual quantum programs is frequently addressed problem in research works \cite{Khatri2019quantumassisted, 2020arXiv200304462Y,9605287,madden2021best,smith2021leap}.
For example, in Ref.~\cite{madden2021best}, published during the preparation of this paper, the authors proposed a strategy to optimize both the single qubit rotations and the gate structure by eliminating trivial single-qubit gates and simplifying the remaining gate structure using $CNOT$ gate identities. 
This method enabled them to achieve significant compression for some of the studied quantum circuits, while for the majority of the addressed circuits the compression turned to be less pronounced.

In this work, we focus on the decomposition of general unitaries by considering random unitary matrices representing the worst-case scenario, and we leave the study of special quantum programs for a future work.
Although we can't provide a rigorous argument to decide whether the SO approach is mathematically exact, our numerical results do not exclude this possibility.
We also tested our algorithm to compile general 3- and 4-qubit quantum programs  for the well known $5$-qubit device QX2 \cite{QX2} of IBM and for the heavy hexagonal lattice design implemented in the falcon quantum processors \cite{Jurcevic_2021} and aimed to be used in forthcoming generations of IBM quantum processors as well.
The decomposition of the addressed quantum programs for these architectures needed only about $13\%$ more $CNOT$ gates than in the ideal full connectivity case. 
The proof-of-principle implementation of our algorithm [called Sequential Quantum Gate Decomposer (SQUANDER)] is accessible via GitHub \cite{SQUANDER_github}.

The rest of the paper is organized as follows. In Sec.~\ref{sec:decsription} we describe the details of the SO algorithm. 
Then in Sec.~\ref{sec:numerical_results} we numerically analyse the performance of the SO algorithm implemented in the SQUANDER package to decompose general $3$ and $4$-qubit unitaries in case of full connectivity design. 
Section \ref{sec:specific_gates} provides a detailed report on how to perform gate decomposition implementing custom quantum gate types and adopt the decomposition process to sparse connectivity designs.
Finally, in Secs.~\ref{sec:squander} and \ref{sec:conclusions} we provide implementation   {details} of the SQUANDER package and summarize the main results of our work, respectively.

\section{Description of the heuristic algorithm} \label{sec:decsription}
 
\begin{figure*}
     \centering
     \includegraphics[width=0.9\textwidth]{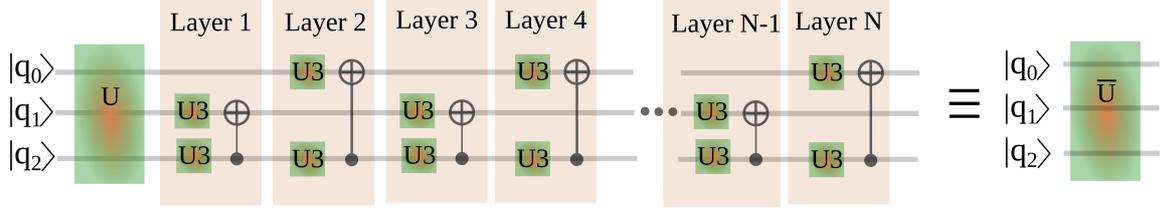}
     \caption{The gate structure composed from $U3$ and $CNOT$ gates used in the decomposition of a three-qubit unitary $U$. The $U3$ gates are represented by green boxes, while the $CNOT$ gates are expressed by vertical lines between two qubits. The gate structure applied after the quantum circuit $U$ disentangle qubit $q_2$ from the qubits $q_0$ and $q_1$. The elementary gates are grouped into layers that are individually optimized one after another until the global minimum of the cost function $f_{sub}$ is not reached up to a given accuracy.
     The unified quantum circuit composed from the unitary $U$ and the operation layers $1$ to $N$ can be formally described by a quantum circuit denoted by $\overline{U}$}
     \label{fig:layers}
 \end{figure*}

The central idea of our heuristic algorithm was inspired by the logic of neural networks, where a pre-constructed structure of function nodes is trained on data sets to mimic some specific function behaviour.
In our case the function nodes in the network are chosen from a universal set of one- and two-qubit gates, which can be considered as linear functions acting on specific qubit states of the quantum register. 
The computing nodes are connected via quantum wires (representing the timeline of the qubits) and via the action of two-qubit gates.
The initially separated qubits become entangled due to the action of a general quantum program represented by a unitary $U$ (see Fig.~\ref{fig:layers}).
Since the final purpose of the gate network is to get the qubits disentangled from each other, the trained gate structure would be the decomposition of $U^{-1}=U^{\dagger}$, and the training procedure would be the process of the decomposition.

A fundamental difference compared to machine learning is, however, that in order to decompose a unitary, we need to   {approach} the global minimum of the cost function very precisely.
Secondly, each unitary $U$ would need an individual optimization process to find a   {suitable} minimum of the cost function (  {that is sufficiently close to the global minimum}), so the optimization process must be very efficient since it needs to be performed every time we want to decompose a quantum program.
Also, the theoretical lower limit of the $CNOT$ gates necessary to decompose general unitaries scales with $\sim 4^n$, so the number of the free parameters associated with the single-qubit gates surrounding the $CNOT$ gates would also increase rapidly with the number of qubits involved in the problem.
In summary, the optimization process needs to be efficient and it should be able to find   {a close-to-global} minimum in a large parameter space at the same time.
These requirements poses a serious challenge on the optimizing algorithm. 
Surprisingly, later in this section we will see that it is possible to design such an algorithm, even if it is specific to our problem, and it highly depends on some peculiar properties of the gates incorporated into the gate design.
Now we turn our attention to the construction of the universal gate structure being later trained to decompose quantum programs.

\subsection{Description of the gate structure} \label{sec:gate_struct}

Our heuristic algorithm follows the strategy to disentangle qubits from the others sequentially, i.e. one by one. 
The disentanglement of a single qubit is achieved by an application of a systematic gate structure on the qubits similarly to the technique reported in \emph{Theorem 9} of Ref.~\cite{1629135} used for state preparation. 
To this end we generalized the concept of multi-controlled gates \cite{PhysRevLett.93.130502} by applying single-qubit rotations to both the target and control qubits, as can be seen in Fig.~\ref{fig:layers} for the case of   {an} $n=3$ qubit register.
By introducing additional degrees of freedom into the system, one might expect that fewer $CNOT$ gates would   {be} necessary to disentangle a qubit from the others.
On the other hand, it also becomes harder to determine the optimized parameters of the $U3$ gates. 
\begin{figure}
    \centering
     \includegraphics[width=0.35\textwidth]{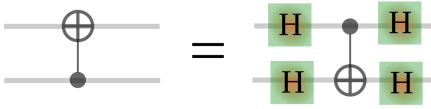}
     \caption{Expressing a $CNOT$ gate in terms of another $CNOT$ gate of reversed orientation and four Hadamard gates acting on both the control and target qubits.}
     \label{fig:CNOT}
 \end{figure}
 Since it is not possible to obtain these parameters by a deterministic procedure we determine them by a recursive optimization process discussed later in Sec.~\ref{subsec:opt}.
 (We note, that a $CNOT$ gate can be expressed using another $CNOT$ gate with reversed orientation and with four Hadamard gates acting on the control and target qubits before and after the reversed $CNOT$ gate, see Fig.\ref{fig:CNOT}. Hence, the orientation of the $CNOT$ gates in Fig.~\ref{fig:layers} would not play an important role.) 
 In order for easier understanding, we group the elementary single- and two-qubit gates into layers according to the pair of qubits on which they apply transformation.
 For example, the first layer in Fig.~\ref{fig:layers} has a direct effect on qubits $q_1$ and $q_2$, while the second layer acts on qubits $q_0$ and $q_2$.
 The generalized gate structure used to disentangle a qubit from the others follows the general logic that (i) each layer consists of two $U3$ gates and one controlled two-qubit gate acting on a pair of qubits; (ii) the gate structure is constructed from periodically repeating sequences of the layers; (iii) each layer must involve the qubit that we want to disentangle from the others; (iv) each of the remaining qubits must be present exactly once in the shortest periodically repeating sequence of the layers.
 
 The rules (i)-(iv) fully determine the number of layers in the shortest period of the gate structure, while the order of the layers inside one period might be arbitrary.
 The total number of the layers in the design is a function of the number of the involved qubits and it also depends on the precision of the decomposition we want to achieve in a given time limit. 
 In Sec.~\ref{subsec:opt}. we will provide detailed numerical analysis of our method and discuss how much of layers (or equivalently $CNOT$ gates) are necessary to disentangle a qubit from the remaining $n-1$ ones.
 
 Finally, we notice that the efficiency of our decomposing algorithm does not depend on the specific orientation of the individual $CNOT$ gates. Thus, we arrange the $CNOT$ gates in the gate design such that the control qubit is always the qubit we are about to disentangle from the others.
 In addition, the $CNOT$ gates in the design can be replaced by other two-qubit controlled gates (although with some limitations to be discussed later), such as $CZ$ (controlled rotation around the $z$ axis) or $CH$ (controlled Hadamard) gates.
 
 \subsection{The cost function of the optimization process}
 
 As we described in previous sections, the goal of the gate structure depicted in Fig.~\ref{fig:layers} (with appropriately chosen parameters of the $U3$ gates) is to disentangle the qubit $q_2$ from the others. 
 The initially separable system of three qubits becomes entangled due to the action of the general three-qubit unitary $U$. 
 The combined action of the unitary $U$ and the decomposing gate structure can be expressed by a "joined" unitary $\overline{U}$, which is the product of $U$ and the other gate operations present in the system.
 In general, the unitary $\overline{U}$ shown at the right side of Fig.\ref{fig:layers}) can be organized in a block form
 \begin{equation}
     \overline{U} = \begin{pmatrix}
     \overline{U}_{00} & \overline{U}_{01} \\ \overline{U}_{10} & \overline{U}_{11}
     \end{pmatrix}\;, \label{eq:block_struct}
 \end{equation}
 where submatrices $\overline{U}_{ij}$ ($i,j\in\{0,1\}$) describe the dynamics of qubits $q_0$ and $q_1$, provided that qubit $q_2$ undergoes a transition from state $|j\rangle$ to state $|i\rangle$.
 If $q_2$ is independent from the rest of the qubits, then submatrices $\overline{U}_{ij}$ differ only by a constant factor from each other, and
 \begin{equation}
     \overline{U}_{ij}\cdot\overline{U}_{pq}^{\dagger} = \kappa_{ij}^{pq}\mathbb{1}\;,  \label{eq:submatrices_prod}
 \end{equation}
 where $\mathbb{1}$ is the identity matrix and $\kappa_{ij}^{pq}$ is the upper left element of the product matrix $\overline{U}_{ij}\cdot\overline{U}_{pq}^{\dagger}$ calculated for indices $i,j,p,q\in\{0,1\}$.
 The cost function $f_{sub}$ quantifying the entanglement of qubit $q_2$ with the rest of the qubits can be then defined as the "distance" of the submatrices $\overline{U}_{ij}$ from the equality given by Eq.~(\ref{eq:submatrices_prod}):
 \begin{equation}
     f_{sub}(\overline{U}) = \sum\limits_{l,m=1}^{2^{(n-1)}}\sum\limits_{ijpq\in(0,1)} \left|\left(\overline{U}_{ij}\overline{U}_{pq}^{\dagger} - \kappa_{ij}^{pq}\mathbb{1}\right)_{lm}\right|^2 \label{eq:sumbmatrix_cost}
 \end{equation}
 In Eq.~(\ref{eq:sumbmatrix_cost}) the indices $l,m$ label the rows and columns of the matrix enclosed by the parentheses. The upper limit of these indices is determined by $n$ labeling the qubit that we are currently disentangling from the others. 
 (In our specific case depicted in Fig.~\ref{fig:layers} $n=2$, since qubit $q_2$ is about to be disentangled from the rest.)
 
 By summing up the squared norm of the matrix elements in Eq.~(\ref{eq:sumbmatrix_cost}) we end up with a single number characterizing the entanglement of qubit $q_n$ with the rest of the qubits.
 The cost function $f_{sub}$ reaches a global minimum $0$ if Eq.~(\ref{eq:submatrices_prod}) holds, i.e. qubit $q_n$ is independent from the other ones. 
 By reaching the global minimum of $f_{sub}$ over the free parameters of the $U3$ gate operations we end up with a gate structure that is tuned to disentangle the qubit $q_n$ from the others.
 
 The described procedure can be applied for an arbitrary number of qubits, however, the total number of layers in the decomposing gate design is expected to increases exponentially with $n$.
 Before we present our numerical results we describe an efficient optimization method to solve the $f_{sub}\rightarrow0$ optimization problem.

\subsection{Solving the optimization problem} \label{subsec:opt}
 
In this section we provide an efficient method to solve the optimization problem $f_{sub}\rightarrow0$. 
The number of gate layers necessary to decompose the unitary $U$ might vary from problem to problem, but depending on the number of qubits involved in the quantum program there is a maximal number of gate layers $N$ involved in the decomposing gate structure that turns to be sufficient to approximate any $N$-qubit unitary $U$ with high numerical accuracy.
However, with the number of the applied layers, the number of free parameters of the optimization problem also increases. 

\begin{figure}
     \centering
     \includegraphics[width=0.4\textwidth]{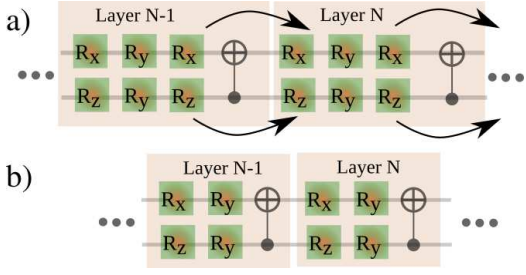}
     \caption{The rotations around $z$ and $x$ axis in subfigure a) can be merged into the successive layer of gate operations. Consequently, each layer can be parametrized with $4$ parameters.}
     \label{fig:4params}
 \end{figure}

According to the reasoning of Ref.\cite{Iten_2016}, rotations around the $z$ and $x$ axis commute with $CNOT$ gates on the control and target qubits, respectively. Since a general single-qubit rotation can be expressed by the combinations $R_zR_yR_z$ and $R_xR_yR_x$, one of the $R_z$/$R_x$ gates can be merged into the successive layer as indicated in Fig.~\ref{fig:4params}.
Consequently, it is sufficient to keep only two free parameters per $U3$ operation, and eventually we end up with $4$ free parameters per layer to be optimized.
Despite of this reduction of the free parameters, the optimization problem $f_{sub}\rightarrow0$ still poses a serious challenge to be solved.
Fortunately, it turns out that the problem can be solved in an iterative way by grouping the layers into blocks that can be individually optimized, while the parameters of the other blocks are kept constant. 
After we found the conditional minimum of the cost function $f_{sub}$ by optimizing the parameters of a given block, we continue the procedure by optimizing the parameters of the following block and we repeat this step over and over again. 
In each iteration step we optimize only a single block of layers and keep the rest of the free parameters constant. 
Following this sequential procedure (also inspiring the naming of the SO algorithm) the cost function of the optimization problem gets smaller in each iteration step until it converges to a global minimum.

Although we can not provide a rigorous mathematical explanation, the outlined block-wise optimization procedure showed a very robust tendency   {in approaching} the global minimum of the cost function $f_{sub}$.
We believe that these exceptional properties of the model are related to the presence of the two-qubit controlled gates in the design.
The sequence of the abrupt transformations (related to the controlled gates) on the initial input state induces large changes in the parameter landscape of the upcoming layers when the parameters of the optimized block are changed.  
Thus, the block-wise optimization of the system has a self-reinforcing tendency to explore large areas in the parameter space favouring to find a spot corresponding to a global minimum.

We tried the same block-wise optimization procedure using parametrized two-qubit operations, such as controlled rotation gates.
In these gates the free parameter is the angle of the controlled rotation making the abrupt state transformation less pronounced.
In this case the state transformation over the sequence of the layers becomes much smoother and the optimization process rather converges to   {a higher} local minimum instead of a global one.

Finally we notice, that the size of the computational blocks can be chosen arbitrary, as far as they remain small enough to be tractable via classical algorithms, such as the Broyden–Fletcher–Goldfarb–Shanno (BFGS) optimization algorithm\cite{kelley1999iterative} used in our implementation. 
Also, since the performance of the outlined algorithm highly depends on the time-scale of a single-block optimization, it is a good practice to keep the size of the blocks low.

\subsection{Decomposition of general unitaries}

\begin{figure*}
     \centering
     \includegraphics[width=0.9\textwidth]{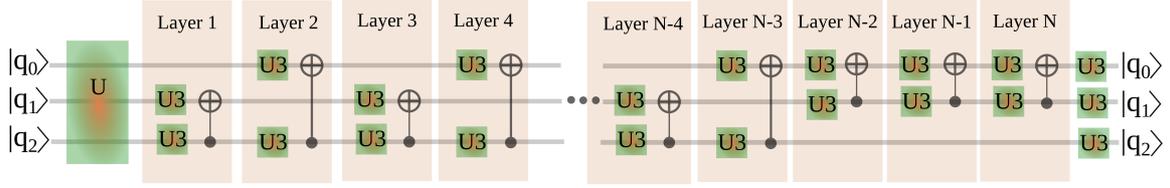}
     \caption{The scheme of the full gate structure reverting the qubits $q_i$ [$i\in(0,1,2)$] into their initial separable state. The qubit $q_2$ is being disentangled from the others by the first $N-3$ layers. The remaining two qubits are disentangled via the operations of layers $N-2$, $N-1$ and $N$. Finally, all of the independent qubits are rotated into their initial state.}
     \label{fig:layers_full}
\end{figure*}

In previous sections we introduced the main ideas to construct a gate design to disentangle a single qubit from the others and formulated the central mathematical problem to be solved to this end.
The outlined procedure can be repeated recursively to disentangle all of the qubits one after another. 
In each recursive step a new unitary $U_{\rm new}=\overline{U}_{ij}$ is chosen from the subblocks of the previously disentangled $\overline{U}$ and the gate structure is extended by new decomposing layers in order to disentangle another qubit from the others.
Since the subblocks $\overline{U}_{ij}$ differ from each other only by a scalar factor, the choice of the subblock is arbitrary. 
Figure \ref{fig:layers_full}. shows the gate structure used to fully decompose a general three-qubit unitary $U$.
The first $N-3$ layers are used to disentangle qubit $q_2$ from the rest and layers $N-2$, $N-1$ and $N$ are responsible to disentangle the remaining two qubits $q_1$ and $q_0$. 
(To decompose a general two-qubit unitary one needs not more than $3$ $CNOT$ gates \cite{PhysRevA.69.032315,PhysRevA.69.062321})
After all the qubits became independent from each other, we can drive them into their initial state by applying individual $U3$ rotation on them. 
  {At the end of the algorithm we perform a final fine tuning on the free parameters of the full decomposing quantum circuit in order to get as close as possible to the synthesised unitary. In this optimisation process we use the metric based on the Frobenius norm defined in Refs.\cite{9605287,madden2021best} to quantify the distance between the initial and the approximated unitary.}
Strictly speaking, the constructed quantum circuit provides the inverse of the initial quantum program $U$ up to a constant phase factor.
Since mathematically the inverse of a unitary is defined by it's complex transpose, the gate structure we have just constructed can be interpreted as the gate decomposition of $U^{\dagger}$.

We notice, that in our implementation of the algorithm we perform the disentanglement of the qubits in a specific order starting with the qubit with the highest label and continue with qubits in descending order of labels. 
In order to perform the decomposition in different order of the qubits one need to reorder the list of the qubits. 
In Sec.~\ref{sec:missing_con} we provide a specific use case where the reordering of the qubits is needed to decompose a unitary for a specific hardware with connection limitations between the qubits.

\section{Numerical results} \label{sec:numerical_results}

In the forthcoming sections we examine the numerical properties of the SO algorithm. 
The most important question is to determine the necessary resources needed to decompose a general quantum program into a sequence of quantum logical gates. 
Before turning our attention to the decomposition of a general unitary, we first examine haw many decomposing layers are needed to disentangle a single qubit from a register of entangled $n$ qubits. 
\begin{figure}
     \centering
     \includegraphics[width=0.45\textwidth]{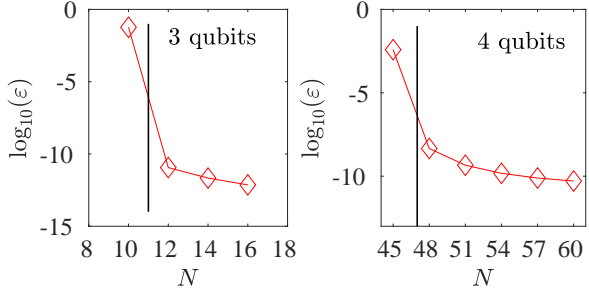}
     \caption{The numerically obtained averaged minimum of the cost function $f_{sub}$ defined as $\varepsilon = \textrm{min}(f_{sub})$ in terms of the number $N$ of the implemented decomposing layers. 
     $\overline{\varepsilon}$ was computed as the average of $10$ independent runs of minimizing $f_{sub}$ with a new random unitary in each run. 
     The vertical black lines represent the theoretical lower bound $N_{theo}(n)-N_{theo}(n-1)$ needed to disentangle the $n$-th qubit from the rest.
     As one can see, $\overline{\varepsilon}$ starts to rapidly increase below the theoretical lower bound, while it saturates above this limit.
     }
     \label{fig:epsilon}
 \end{figure}
Although our algorithm is not limited to specific two-qubit gates, we start the discussion of our numerical results obtained when $CNOT$ gates were implemented in the gate structure.
The analytic expression $N_{theo}(n)$ provides a starting point to determine the number of layers needed to disentangle a qubit from the others.
In particular, we estimate the minimal number of $CNOT$ gates to $N_{theo}(n)-N_{theo}(n-1)$.
(The decomposition of an $n$-qubit unitary needs at least $N_{theo}(n)$ $CNOT$ gates, which can be split between the disentanglement of the first qubit from the others and the decomposition of the remaining $(n-1)$-qubit unitary needing $N_{theo}(n-1)$ $CNOT$ gates.)
Here we examine how close we can optimize the parameters of $f_{sub}$ to obtain the target minimum of $0$ needed to disentangle a qubit from the rest. 
If the number of the layers (which is equal to the number of two-qubit controlled gates) is larger than the theoretical lower limit, we expect that the free parameters of $f_{sub}$ can be optimized to come close to the $f_{sub}\rightarrow0$ solution with some numerical precision $\varepsilon$.
As soon as we decrease the number of the layers below the theoretical lower limit, the global minimum of $f_{sub}$ should rise above $0$, since there would not be sufficient $CNOT$ gates in the circuit to meet the desired solution.
Figure \ref{fig:epsilon}. shows the average value of $\varepsilon = \textrm{min}(f_{sub})$ in terms of the applied number $N$ of   {the} decomposing layers.
As we see in the figure, $\varepsilon$ picks up large values below the theoretical lower limit indicated by black vertical lines. 
In this regime the optimization problem $f_{sub}\rightarrow0$ has no solution.
Above the theoretical limit the solution of the optimization problem saturates on $\varepsilon\leq \mathcal{O}(10^{-8})$ which can not be further decreased due to accumulated numerical errors during the calculations.
The numerical precision of the calculations is limited by the double precision format of floating point number representation implemented in BLAS libraries\cite{blackford2002updated}. 
Since matrix multiplication is the most intensively used operation during the optimization process, it has the greatest influence on the final numerical error.
In principle, it would be possible to implement matrix multiplication using higher precision number representation in exchange of rapid increase of the execution time.
However, the $64$ bit architecture of modern CPU-s poses hardware limitation to do efficient matrix multiplication using higher precision number format.
(We notice, that CPUs still provide floating point operation unit for extended precision numbers with $80$ bit registers, but these hardware units do not support vectorization of floating point operations.)

At the end, one needs to make a trade-off between the final precision of the decomposition and the computational time. 
Our numerical analysis showed that double precision floating point number format is sufficient to achieve $\lVert U - U_{approx}\rVert_2\sim\mathcal{O}(10^{-5})$ precision in the decomposition of general $3$, $4$ and $5$-qubit unitaries on a reasonable time-scale, while the number of implemented $CNOT$ gates is very close to the theoretical lower limit.
Here $U$ is the initial unitary and $U_{approx}$ is the quantum program standing for the decomposed quantum circuit.
The norm $\lVert A\rVert_2$ stands for the spectral norm of matrix $A$, which corresponds to the largest singular value of $A$, i.e., the square root of the largest eigenvalue of the matrix $A^\dagger A$, where $A^{\dagger}$ denotes the conjugate transpose of $A$.
(We note that $U_{approx}$ might differ by a global phase factor from the initial unitary $U$. This global phase factor should be removed when calculating the error of the decomposition.)

\begin{figure}
     \centering
     \includegraphics[width=0.45\textwidth]{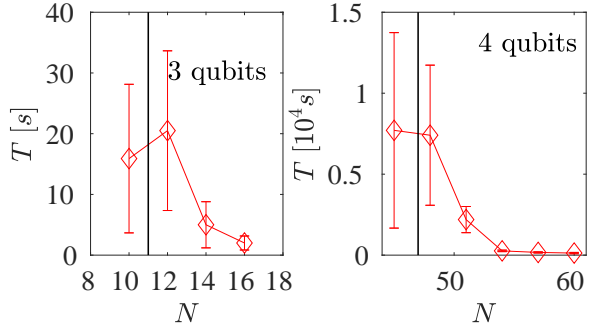}
     \caption{The execution time to find the global minimum of the cost function $f_{sub}$. 
     The execution time was measured by averaging $10$ independent runs to disentangle a qubit from the others as a function of the implemented layers. 
     The standard deviance of the measured execution times is represented by the errorbars.
     The vertical black lines represent the theoretical lower bound $N_{theo}(n)-N_{theo}(n-1)$ needed to disentangle the $n$-th qubit from the rest.
     The numerical simulations were done on a traditional desktop PC with \emph{Intel Core i7} CPU and with $8$ GB of RAM.
     }
     \label{fig:runtime}
 \end{figure}

The execution time necessary to find the global minimum of $f_{sub}$ highly depends on the number $N$ of the implemented layers in the circuit.
Figure \ref{fig:runtime}. shows the execution time of the optimization process in terms of the applied number $N$ of decomposing layers.
As one can see the execution time highly increases as the number of layers comes close to the theoretical lower limit indicated by the black vertical lines. 
By increasing the number of the free parameters in the problem by adding extra decomposing layers, the convergence to the global minimum of $0$ becomes much faster, so it becomes computationally easier to decompose a unitary by increasing the number of $CNOT$ gates.  
Another option to reduce the execution time is to set a hard limit on $\varepsilon$, and when the optimization process reaches the limit $\varepsilon_0$, the implementation of the algorithm steps forward to disentangle the next qubit.
By setting an optimization limit $\varepsilon_0$ to $10^{-8}$ the final error of the decomposition is turned to be as small as $\lVert U - U_{approx}\rVert_2\sim\mathcal{O}(10^{-5})$   {and $\mathcal{O}(10^{-4})$} for a general $3$ and $4$-qubit unitary,   {respectively}.
At the same time, the execution time of the decomposition decreases to several seconds for $3$-qubit unitary incorporating 15 $CNOT$ gates in total and to $2$ minutes in case of a $4$-qubit unitary incorporating $75$ $CNOT$ gates in total.
(The execution time was measured on an \emph{Intel Core i7} CPU having $8$ threads.)
However, these execution times are still not competitive with current implementations of the CSD or QSD implementations\cite{krol2021efficient}, for larger problems than $4$ qubits, the decomposition time exceeds the limit of an acceptable time scale.
Still, the low $CNOT$ gate count close to the theoretical lower limit might justify the increased execution time in many use-cases. 
For example, as we will see in Secs.~\ref{sec:specific_gates} and \ref{sec:missing_con}, the SO algorithm is also applicable to optimize a gate structure on architectures with specific properties or limitations.
In addition, the developed SO algorithm can be mixed with QSD to optimize the number of gates by letting the SO algorithm to decompose small unitaries generated during the QSD procedure\cite{krol2021efficient}.

\begin{table*}[ht!]
\centering
\begin{tabular}{||c c c c c c||} 
 \hline
  Number of qubits & 2 & 3 & 4 & 5 & $n$\\ [0.5ex] 
 \hline\hline
 Iterative unentangling \cite{1629135} & 8 & 62 & 344 & 1642 & $2\cdot4^n - (2n+3)\cdot2^n + 2n$ \\
 Givens rotations \cite{PhysRevLett.92.177902,krol2021efficient} & 4 & 64 & 536 & 4156 & $\approx 8\cdot4\cdot4^n$ \\
 Recursive CSD \cite{tucci} & 14 & 92 & 504 & 2544 & $\frac{1}{2}\cdot n \cdot 4^n - \frac{1}{2}\cdot2^n$ \\
 Recursive CSD (optimized) \cite{mottonen2005} & 4 & 26 & 118 & 494 & $\frac{1}{2}\cdot 4^n - \frac{1}{2}\cdot2^n - 2$ \\
 QSD \cite{1629135,krol2021efficient} & 6 & 36 & 168 & 720 & $\frac{3}{4}\cdot4^n - \frac{3}{2}\cdot2^n$ \\
 QSD (optimized)\cite{1629135} & 3 & 20 & 100 & 444 & $(23/48)\cdot(4^n) - \frac{3}{2}\cdot2^n + \frac{4}{3}$ \\ 
 \textbf{Sequential optimization\footnote{The numbers given in the table represent the limiting case of the sequential optimization algorithm resulting in the highest execution time. The decomposition would be faster by orders when more $CNOT$ gates are used for the decomposition. For further details see the main text.}} & \textbf{3} & \textbf{15} & \textbf{63} & \textbf{267} & Not available \\
 Theoretical Lower Bounds\cite{PhysRevA.69.062321} & 3 & 14 & 61 & 252 & $\frac{1}{4}\left(4^n - 3n - 1\right)$ \\ [1ex] 
 \hline
\end{tabular}
\caption{The number of $CNOT$ gates needed to decompose a general unitary for the individual algorithms. The sequential optimization implemented in the SQUANDER package provides the lowest number of $CNOT$ gates needed for the decomposition   {$\mathcal{O}(10^{-5})$ and $\mathcal{O}(10^{-4})$ for $3$ and $4-5$ qubit unitaries, calculated with the spectral norm metric defined in Sec.~\ref{sec:numerical_results}}.}
\label{table:CNOT_gates}
\end{table*}
In Table \ref{table:CNOT_gates} we summarize our numerical results on unitary decomposition by reporting the lower bound of $CNOT$ gates needed to decompose a general unitary and we compare our results to other algorithms. 
Our studies revealed, that any three-qubit unitary can be approximated using at most $15$ $CNOT$ gates by the SO algorithm.
In addition, $63$ and $267$ $CNOT$ gates are sufficient to decompose a $4$ and $5$-qubit unitary, respectively up to precision of   {$\mathcal{O}(10^{-5})$ and $\mathcal{O}(10^{-4})$ for $3$ and $4-5$ qubit unitaries, calculated with the spectral norm metric defined earlier in the text}.

\section{Architecture specific decomposition} \label{sec:specific_gates}

The most challenging hardware components of quantum computers are the two-qubit controlled gates that are used to induce entanglement between the qubits. 
Depending on the underlying architecture, the nature of the interaction between the qubits might suit for the creation of different kinds of two-qubit gates.
For example, in Ref. \cite{PhysRevLett.107.080502} a microwave irradiation could be used to form a cross resonant gate between the qubits, Refs.~\cite{Barends2014,PhysRevApplied.15.064005} reported on experimental demonstration of controlled-phase gates between transmon qubits, while another realization of controlled-phase gates between two photonic qubits encoded in photonic fields stored in cavities or between time-bin qubits were accomplished in Refs.~\cite{PhysRevLett.124.120501} and \cite{PhysRevApplied.13.034013}, respectively. 
Usually, the $CNOT$ gates in these platforms are generated as a composite gate operation formed from one two-qubit contolled gate and from suitable one-qubit rotations\cite{PhysRevApplied.15.064005, PhysRevApplied.13.034013}.

Another typical issue on quantum hardware is related to limitations in inter-qubit connectivity.
Since two-qubit gates can be in many experimental situations realized only between qubits close to each other a gate decomposition of a quantum program needs to be adopted to these limitations.
In most cases this is accomplished by adding extra $SWAP$ gates into the circuit, which tends to highly increase the gate cunt.
However, in order to increase the total fidelity of a quantum circuit, it would be more suitable to develop a decomposition algorithm that natively supports the limitations of the targeted quantum hardware. In this section we show that the SO algorithm can be adopted to such needs.

\subsection{Custom two-qubit controlled gates in the decomposition} \label{sec:mixing_gates}
 
As we mentioned earlier in Sec.~\ref{subsec:opt}, the SO algorithm is not limited to specific choice of the gates used in the decomposition. 
The decomposition can be executed with arbitrary, but parameter-less two-qubit controlled gates as well.

\begin{figure}
     \centering
     \includegraphics[width=0.35\textwidth]{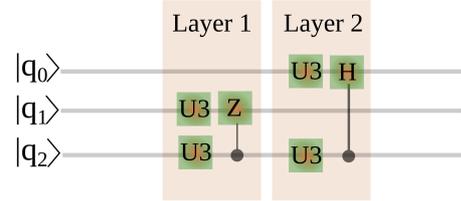}
     \caption{A period of a custum gate structure used to disentangle qubit $2$ from qubits $0$ and $1$ in a three-qubit problem. Qubit $2$ is connected to qubit $1$ via controlled pahse gate, while the interaction with qubit $0$ is realized via controlled Hadamard gate. The source code to generate such gate structure is described in Alg.~\ref{alg:custom_gates}.
     }
     \label{fig:custom_gates}
 \end{figure}

The SQUANDER package (for details see Sec.~\ref{sec:squander}.) provides a Python (and native C++) interface to construct custom gate structure for the decomposition.
\begin{algorithm*}[!ht]
\SetAlgoLined
\LinesNumbered
\SetKwFunction{FCustomGates}{CustomGates}

\SetKwInOut{Input}{input}\SetKwInOut{Output}{output}
\SetKwFor{For}{for}{:}{}
\SetKwIF{If}{ElseIf}{Else}{if}{:}{elif}{else:}{}

\Output{A period in the three-qubit gate structure used to disentangle qubit $q2$ from the rest.}
\SetKwProg{Fn}{def}{:}{}
\textbf{from} qgd\_python.gates.qgd\_Operation\_Block \textbf{import} qgd\_Operation\_Block \\\hspace{0.1cm}

\Fn{\FCustomGates{}}{

    \# The number of the qubits in the register \\
    qubit\_num = 3 \\\hspace{0.1cm}
    
    \# creating an instance of class representing a period of the gate structure \\
    Gate\_Period = qgd\_Gates\_Block( qbit\_num ) \\\hspace{0.1cm}
    
    \# create the first decomposing layer with $CZ$ gate \\
    Layer1 = qgd\_Gates\_Block( qbit\_num ) \\\hspace{0.1cm}

    \# add U3 gates acting on qubits 0 and 1 with two free parameters ($\Theta$ and $\lambda$) to Layer1 \\
    Theta = True \\
    Phi = False \\
    Lambda = True  \\     
    Layer1.add\_U3( 1, Theta, Phi, Lambda )   \\        
    Layer1.add\_U3( 2, Theta, Phi, Lambda )   \\\hspace{0.1cm}
    
    \# add $CZ$ gate to layer1 with control qubit 2 and target qubit 0\\
    Layer1.add\_CZ( 1, 2) \\ \hspace{0.1cm}
    
    \# create the second decomposing layer with controlled Hadamard gate \\
    Layer2 = qgd\_Gates\_Block( qbit\_num ) \\\hspace{0.1cm}

    \# add U3 gates acting on qubits 1 and 2 with two free parameters ($\Theta$ and $\lambda$) to Layer2 \\
    Theta = True \\
    Phi = False \\
    Lambda = True  \\     
    Layer2.add\_U3( 0, Theta, Phi, Lambda )   \\        
    Layer2.add\_U3( 2, Theta, Phi, Lambda )   \\\hspace{0.1cm}
    
    \# add controlled Hadamard gate to layer2 with control qubit 2 and target qubit 1\\
    Layer2.add\_CH( 0, 2) \\\hspace{0.1cm}
    
    \# add the decomposing layers to the period of the decomposing gate structure \\
    Gate\_Period.add\_Gates\_Block( Layer1 ) \\
    Gate\_Period.add\_Gates\_Block( Layer2 ) 
    
    return Gate\_Period
    
    }
    
 \caption{Source code example using the Python interface of the SQUANDER package to construct a period of decomposing gate structure with custom two-qubit gates in the design.} \label{alg:custom_gates}
\end{algorithm*}
In Alg.~\ref{alg:custom_gates}. we provide a brief overview on how to define such custom gate structure to disentangle qubit $q2$ form qubits $q0$ and $q1$ in a three-qubit problem.
It is sufficient to define only a single period of the gate structure, since this period is repeated during the decomposition automatically. 
(The number of periods in the decomposing gate design can be set by an individual parameter. 
For details see the tutorial material of the SQUANDER package\cite{SQUANDER_docs}.)
First we import the necessary package from SQUANDER and define the number of the qubits by a local variable. 
Then we create an instance of class \emph{qgd\_Gates\_Block} representing one period in the decomposing gate structure. 
The individual layers \emph{Layer1} and \emph{Layer2} of the gate design are represented by the same type of class as the period itself. 
Finally we add the two- and one-qubit gates to the layers using the methods implemented in the class \emph{qgd\_Gates\_Block}. 
We chose to add controlled phase ($CZ$) gate to the first layer and controlled Hadamard gate to the second layer.
One period of the constructed gate structure is shown in Fig.\ref{fig:custom_gates}.
 Equivalently to the case when only $CNOT$ gates were implemented in the design, there is needed at least $12$ decomposing layers to disentangle qubit $q2$ from the others. Since the implemented two-qubit gates are related to the $CNOT$ gate by one-qubit rotations, this equivalence is straightforward.
 
 The SQUANDER package contains a working example implementing the described use-case, it can be also found in the tutorial material at \cite{SQUANDER_docs}. 
 We notice, that the code example in Alg.~\ref{alg:custom_gates}.   {shows} only the steps to construct custom gate structure, to implement it in actual   {gate synthesis} additional steps are needed that are described in the tutorial material as well.

\subsection{Adopting decomposition to a hardware specific connectivity structure} \label{sec:missing_con}

\begin{figure}
     \centering
     \includegraphics[width=0.45\textwidth]{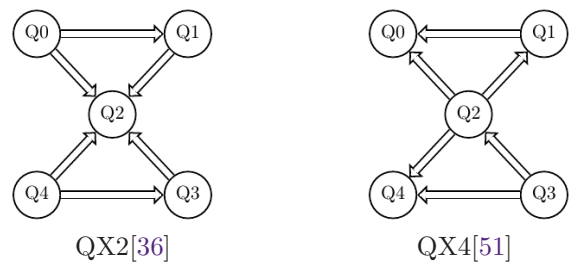} \\
     QX2\cite{QX2} \hspace{3cm} QX4\cite{QX4}
     \caption{Connectivity graph between the qubits in the QX2 and QX5 five-qubit IBM quantum computers \cite{Sisodia2018CircuitOF}. A qubit  shown  at  the  tail  of  the  arrows  can  only work as control qubit and the qubit shown at the head of the arrow can only work as target qubit. 
     }
     \label{fig:IBM5}
 \end{figure}

All numerical results presented in previous sections were obtained for designs without any limitations regarding the connections between the individual qubits, i.e. each of the qubit pairs could be connected by two-qubit gates.
However, in realistic experimental setups only neighbouring qubits can be connected in a direct gate operation. 
In many cases even the orientation of the two-qubit controlled gates are fixed, it is not possible to chose the control or the target qubits on demand.
Examples of such architecture are represented in Fig.~\ref{fig:IBM5} showing the case of the 5-qubit quantum devices of IBM publicly available via IBM Cloud services\cite{IBMcloud}.
The ability to decompose a quantum program into gate structures with sparse connectivity is undoubtedly an important issue of quantum programming.
One theoretically possible way to overcome this issue is to apply swap gates to get distant qubits close to each other where the desired two-qubit controlled operation can be applied on them, and then move the qubits back to their initial position by another swap operations.
In order to increase the fidelity of a quantum circuit constructed this way, there have been proposed several optimization approaches \cite{8491885, Sisodia2018CircuitOF} that might considerably reduce the number of gate operations in   {a} quantum program.
Efficient gate reduction techniques are becoming even more important as the number of qubits continuously increase on the hardware side, while the connectivity density stays low. 
On the $8$-th of August, 2021, IBM announced to implement their future quantum devices basing on the so-called heavy hexagonal lattice design, which offers even less connection between the qubits than previous architectures.

Here in this section we report on a novel approach to optimize a quantum circuit of general unitaries. 
The key point of our approach is to find the best decomposition of a quantum program from the start, without implementing additional swap gates into the decomposed circuit.
To this end we use the SO method on custom gate structure adopted to the architecture of the target hardware.
Our method gives the lowest possible number of $CNOT$ gates in the decomposition of a general unitary, close to the theoretical lower limit.
As an example we show how to decompose a $4$-qubit quantum program on the QX2 $5$-qubit device of IBM\cite{QX2}.

The choice of the four qubits might be in principle arbitrary, as long as the chosen set contains the central $Q_2$ qubit.
In our example we decompose a general $4$-qubit unitary on qubits $Q_0$, $Q_1$, $Q_2$ and $Q_3$ of the QX2.
At the beginning of the decomposition we need to decide the order in which we are going to disentangle the qubits from the others. 
In general, qubits having the most connections are the easiest to disentangle from the others.
However, since qubit $Q_2$ plays a central role in the design, we need to choose another qubit to begin with, otherwise there would be left no direct connection to qubit $Q_3$ and it would be not possible to finish the decomposition without using swap gates. 
For example, we might choose qubit $Q_0$ as the first one to disentangle. (We notice that at this point qubit $Q_1$ would be equivalently good choice.)
Then we can continue the decomposition with getting qubit $Q_1$ independent from the others, and finally we disentangle the remaining two qubits $Q_2$ and $Q_3$.
Since the SQUANDER package process the decomposition in specific order of the qubits (always disentangling the qubit with the highest index), we need to re-label the qubits, so SQUANDER can  disentangle the qubits in the correct order. 
\begin{figure}
     \centering
     \includegraphics[width=0.45\textwidth]{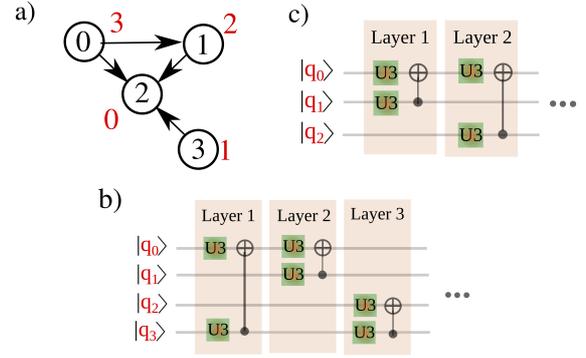}
     \caption{a) Re-labeling the qubits of the original QX2 architecture to perform a decomposition in correct order. The new labels used by the SQUANDER package are indicated by the red numbers, while the black numbers corrspond to the original labeling of the qubits.
          b) and c) The periods of the gate structures to disentangle qubits $3$ and $2$ from the rest, respectively. Here the qubits follows the labeling indicated by the red numbers in a). 
     }  \label{fig:QX2}
\end{figure}

\begin{figure}
     \centering
     \includegraphics[width=0.45\textwidth]{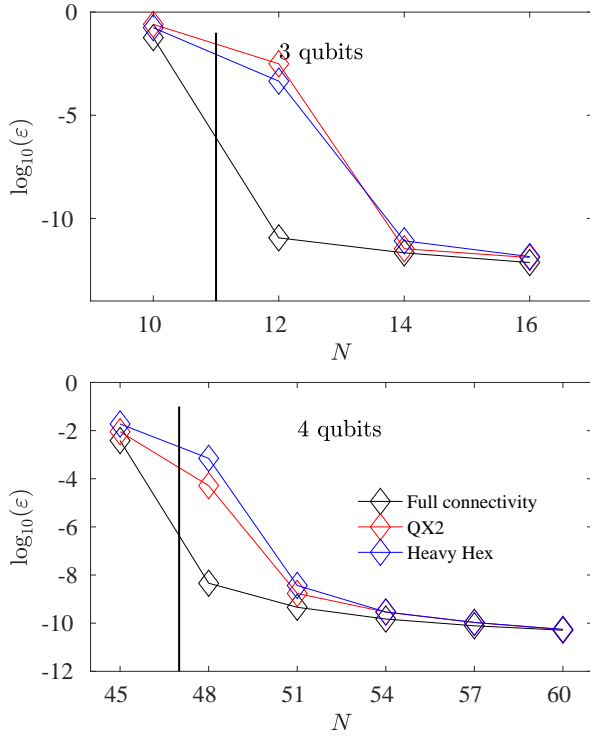}
     \caption{The numerically obtained average minimum of the cost function $f_{sub}$ defined as $\varepsilon = \textrm{min}(f_{sub})$ in terms of the number $N$ of the implemented decomposing layers defined by the design in Figs.~\ref{fig:QX2} and \ref{fig:heavy_hex} for QX2 and heavy hexagonal architectures, respectively.
     $\overline{\varepsilon}$ was computed as the average of $10$ independent runs of minimizing $f_{sub}$ with a new random unitary in each run. 
     The standard deviance of the calculated mean is smaller than the size of the data markers.
     The vertical black lines represent the theoretical lower bound $N_{theo}(n)-N_{theo}(n-1)$ needed to disentangle the $n$-th qubit from the rest.
     As one can see, the reduced connectivity between the qubits in QX2 and heavy hexagonal architectures implies more decomposing layers needed to disentangle a qubit.
     }  \label{fig:QX2_results}
\end{figure}

Figure \ref{fig:QX2}.a) shows the relabeled qubits used in the decomposition. 
In order to construct a suitable gate design to disentangle qubits $q3$ and $q2$, we need to abandon rule (iii) formulated in Sec.~\ref{sec:gate_struct} for the ideal full connectivity case.
Since qubit $q3$ does not have direct connection to all the remaining qubits, it has no point to involve this qubit in all the decomposing layers when disentangling it. 
Instead, as shown in Fig.~\ref{fig:QX2}.b), we mediate the interaction between qubits $q3$ and $q1$ (which are not directly connected) via qubit $q0$, since $q0$ is connected to both $q3$ and $q1$.
Similarly, when it comes to disentangle qubit $q2$, we need to use again qubit $q0$ to mediate between $q2$ and $q1$, as shown in Fig.~\ref{fig:QX2}.c).
Due to these mediated interactions between the qubits one might expect that more $CNOT$ gates would be needed for the decomposition as in the case of ideal full connectivity.
Unfortunately, there is no rule to tell how many decomposing layers would be needed, we know only the lower limit of the $CNOT$ gates from Table.~\ref{table:CNOT_gates}, which is a good starting point to begin with.
Our numerical results shown in Fig.~\ref{fig:QX2_results} indicate that for the QX2 architecture we indeed need more decomposing layers to disentangle qubits from the rest. 
Surprisingly, the precision $\varepsilon$ of the disentanglement comes very close to the full connectivity case by adding only several extra decomposing layers to the circuit.
To disentangle the $4$-th qubit in a four-qubit register one needs $54$ layers (i.e. $54$ $CNOT$ gates) to achieve the same precision as in the full connectivity case, while in the three-qubit case $14$ layers are sufficient to get the same precision.
In overall, the SO algorithm needs at least $17$ $CNOT$ gates to decompose a general three-qubit unitary on the QX2 architecture and achieving the same precision as in the full connectivity case, while $71$ $CNOT$ gates are sufficient in the $4$-qubit case.

In heavy hexagonal structure of the qubits the connectivity between the qubits is even more reduced in order to increase the scalability of the architecture for fabrication.
\begin{figure}
     \centering
     \includegraphics[width=0.45\textwidth]{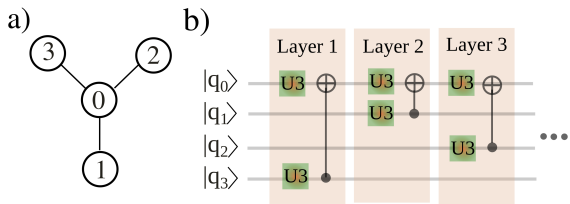}
     \caption{a) Selecting a $4$ qubit sub-system in a heavy hexagonal lattice for a $4$-qubit qunatum program. b) A period of a gate structure to disentangle qubit $3$ from the others.}  \label{fig:heavy_hex}
\end{figure}
Here each unit cell of the lattice consists of a hexagonal arrangement of qubits, with an additional qubit on each edge. 
IBM first introduced the so-called "heavy hex" topology in 2020, the family of falcon quantum processors already implemented this design\cite{Jurcevic_2021}.
According to IBM the key idea behind reducing the number of connections between the qubits is justified by minimizing both the qubit frequency collisions and spectator qubit errors.
However, less connectivity makes quantum circuits harder to implement.
Figure \ref{fig:heavy_hex}.a) shows a $4$-qubit "piece" of the heavy hex structure. 
Even if the $CNOT$ gates are (in principle) bidirectional in these designs, the missing connection between qubits $3$ and $2$ might increase the depth of quantum circuits compared to the QX2 architecture studied previously.
Surprisingly, our numerical results showed that by using the gate structure depicted in Fig.~\ref{fig:heavy_hex}.b) to disentangle qubit $3$ from the others, there is no need for more $CNOT$ gates as were used for the QX2 architecture.
According to Fig. \ref{fig:QX2_results}, the   {obtained} minima of $f_{sub}$ are very close to each other for the QX2 and for the heavy hexagonal architectures, indicating that the missing connection between qubits $2$ and $3$ does not play a crucial role in the decomposition.
It turns out that to decompose a general $4$-qubit (or $3$-qubit) unitary one needs the same number of $CNOT$ gates on both the QX2 and the heavy hexagonal architectures. Comparing this with the case of a fully connected topology, it is surprising how small the CNOT-count overhead is. It would be very interesting to study how this overhead would change for the heavy hexagonal lattice  when we increase the number of qubits.

\section{SQUANDER: implementation of the sequential optimization algorithm} \label{sec:squander}

Generally, a numerical procedure to solve an optimization problem is computationally more demanding than a deterministic or an iterative algorithm used to solve linear algebraic problems.
For this reason, our implementation is not expected to be competitive in speed with other well known decomposition implementations, such as was reported in Ref.~\cite{krol2021efficient}.

Still, in order to increase the computational performance of our implementation of the sequential optimization algorithm the solver engine was implemented in native C++/C programming language.
In principle the execution time of the code can be further decreased by combining dense and sparse matrix operations during the optimization process.
In order to keep the benefits coming with the flexibility and the popularity of a high level Python API we also designed a Python interface for our implementation, enabling to export the results of the decomposition into the well known quantum programming package Qiskit. 
(Interfacing with other quantum packages are under development.)

Our implementation, the Sequential Quantum Gate Decomposer (SQUANDER) package is accessible through a public GitHub repository \cite{SQUANDER_github}, while a documentation of the package is hosted at \emph{CodeDocs[xyz]} site \cite{SQUANDER_docs}.
The SQUANDER package is equipped with the Threading Building Block library \cite{ProTBB} providing an efficient task oriented parallel programming model to achieve an optimal workload balance among the accessible execution units of the underlying hardware avoiding any over-subscription of the resources. 
The register level parallelism via portable SIMD instruction are provided by the incorporation of low level BLAS functions.

\section{Conclusion and Outlook} \label{sec:conclusions}

In this work, we constructed and numerically analyzed an optimization method to approximate general unitaries by quantum circuit made of one- and two-qubit gate operations.
At this point our approach is based on numerical heuristics, it is hard to make any mathematically rigorous arguments about the limits of the algorithm. 
Solving the optimization problem by a sequential technique enables one to get close to the global minimum of the cost function $f_{sub}$ up to an accuracy hitting the numerical precision accumulated by the floating point operations.
The approach works for arbitrary combination of two-qubit controlled gates and gives the lowest number of the two-qubit controlled gates necessary to decompose a general unitary. (For details see Table.~\ref{table:CNOT_gates}.)
However, it is still an open question whether the developed numerical approach can be considered to be an exact solution or not.

We also examined the possibility of using our approach to decompose quantum programs on architectures having sparse connectivity between the qubits. 
Our method natively supports architecture specific features of the underlying architecture.
The decomposition methods like the CSD or QSD,   {on the other hand}, assume ideal,   {all-to-all} connectivity between the qubits. In order to adopt a decomposed quantum circuit to a specific architecture, one needs to include additional swap gates and perform post optimization strategies to reduce the depth of the resulting quantum circuit. 
In the case of the sequential optimization algorithm, however, the decomposition can be adopted to hardware specifics from the start, resulting in a significantly lower depth of the quantum circuit.

Unfortunately, the increased execution time needed to find the optimized parameters of the gate structure poses a serious limitation on the applicability of our algorithm. 
The decomposition of a $7$-qubit unitary takes more than a week, which limits the use-cases of the sequential decomposition algorithm to low number of qubits.
In order to make use of the sequential optimization algorithm in the decomposition of larger unitaries, our algorithm can be combined with other decomposing approaches. 
In particular, QSD follows a similar strategy than our algorithm. 
By disentangling qubits one after another, in each iterative step a new unitary is produced with a less qubits than in the previous iteration step. 
By applying our algorithm at the end of this procedure -- where $3$, $4$ or $5$-qubit unitaries needs to be processed -- one might significantly reduce the number of gates in the decomposition and make steps to adopt the decomposed quantum program to the connectivity specifics of the underlying hardware at the same time.
However, we leave the study of this opportunity for future work.
We believe that our results might trigger new strategies to turn quantum programs into efficient quantum circuits.

\section{Acknowledgements}

The research was supported by the Ministry of Innovation and Technology and the National Research, Development and Innovation
Office within the Quantum Information National Laboratory of Hungary, and was
also supported by NKFIH through the Quantum Technology National Excellence Program
(No.2017-1.2.1-NKP-2017-00001) and  Grants No. 2020-2.1.1-ED-2021-00179, K124152, FK135220, KH129601.
We acknowledge the computational resources provided by the Wigner Scientific Computational Laboratory (WSCLAB) (the formerWigner GPU Laboratory).

\bibliographystyle{IEEEtran}
\bibliography{references}

\end{document}